\documentclass[12pt]{article}
\pdfoutput=1

\usepackage[utf8]{inputenc}

\usepackage{amsfonts}
\usepackage{amssymb}

\usepackage{physics}
\usepackage{tensor} 
\usepackage{subcaption} 

\usepackage[
      colorlinks=true,
      linkcolor=blue,
      urlcolor=blue,
      filecolor=black,
      citecolor=red,
      pdfstartview=FitV,
      pdftitle={},
        pdfauthor={ Michael Gutperle, Kevin Chen},
        pdfsubject={},
        pdfkeywords={},
        pdfpagemode={},
        bookmarksopen=true
      ]{hyperref}

\hypersetup{linktocpage}

\usepackage{color}
\usepackage{graphicx}

\usepackage{sectsty}

\renewcommand{\thefootnote}{\fnsymbol{footnote}}
\renewcommand{\thanks}[1]{\footnote{#1}}
\newcommand{\starttext}{
\setcounter{footnote}{0}
\renewcommand{\thefootnote}{\arabic{footnote}}}
\newcommand{\bea}{\begin{eqnarray}}
\newcommand{\eea}{\end{eqnarray}}
\newcommand{\be}{\begin{equation}}
\newcommand{\ee}{\end{equation}}
\renewcommand\({\begin{equation}}
\renewcommand\){\end{equation}}

\newcommand{\no}{\nonumber}

\numberwithin{equation}{section}

\def\qcq{\quad\quad,\quad\quad}

\usepackage{ dsfont } 

\DeclareMathOperator{\sign}{sign}

\def\no{\nonumber}

\long\def\symbolfootnote[#1]#2{\begingroup%
\def\thefootnote{\fnsymbol{footnote}}\footnote[#1]{#2}\endgroup}

\begin{document}
\setlength{\baselineskip}{18pt}

\starttext
\setcounter{footnote}{0}

\begin{flushright}
January 30, 2019
\end{flushright}

\bigskip

\begin{center}

{\Large \bf Relating AdS$_6$ solutions in type IIB supergravity}

\vskip 0.4in

{\large  Kevin Chen and Michael Gutperle}

\vskip .2in

{ \it Mani L. Bhaumik Institute for Theoretical Physics }\\
{ \it Department of Physics and Astronomy }\\
{\it University of California, Los Angeles, CA 90095, USA}\\[0.5cm]

\bigskip
\href{mailto:kevinchen@physics.ucla.edu}{\texttt{kevinchen@physics.ucla.edu}}\texttt{, }
\href{mailto:gutperle@physics.ucla.edu}{\texttt{gutperle@physics.ucla.edu}}

\bigskip

\bigskip

\end{center}
 
\begin{abstract}

\setlength{\baselineskip}{18pt}

In this note we show that the IIB supergravity solutions of the form AdS$_6\times M_4$ found by Apruzzi et al.~in \cite{Apruzzi:2014qva} are related to the local solutions found by D'Hoker et al.~in \cite{DHoker:2016ujz}. We also discuss how the  global regular solutions found in \cite{DHoker:2016ysh,DHoker:2017mds}  are mapped to the parameterization of  \cite{Apruzzi:2014qva}.

\end{abstract}

\setcounter{equation}{0}
\setcounter{footnote}{0}

\newpage

\section{Introduction}
\label{sec-1}

Five dimensional superconformal field theories take an interesting place among conformal field theories.  They realize  a unique superconformal algebra $F(4)$, they are  strongly coupled in the UV, and many exhibit unusual properties such as enhanced 
exceptional flavour symmetries \cite{Seiberg:1996bd,Morrison:1996xf,Intriligator:1997pq}. Holography is a useful method to study 
strongly coupled CFTs.  However, until recently very few supergravity solutions in ten or eleven  dimensions   dual to five dimensional 
SCFTs  were known.  The first solutions \cite{Brandhuber:1999np,Bergman:2012kr,Passias:2012vp}  were constructed in massive IIA 
supergravity. Special examples of type IIB solutions were constructed from the type IIA  solution using (non-Abelian) T-duality in \cite{Lozano:2012au,Lozano:2013oma}. In \cite{Apruzzi:2014qva} type IIB supergravity solutions were constructed from first principles. The 
solutions take the form of a fibration of AdS$_6$ over a four dimensional base manifold and pure spinor geometry is used to  determine 
the conditions for sixteen unbroken supersymmetries. It was found that the manifold $M_4$ is a $S^2$ fibration over a two dimensional 
space and the problem is reduced to solving two partial differential equations  on this  two dimensional space. In \cite{DHoker:2016ujz} a different approach 
utilizing Killing spinors on an AdS$_6\times S_2$ fibration over a two dimensional Riemann surface $\Sigma_2$  was used  to reduce the 
BPS equations of the bosonic background. It was shown that  local solutions can be expressed in terms of two holomorphic functions 
on the Riemann surface  $\Sigma_2$. Later, regular global solutions were constructed \cite{DHoker:2016ysh,DHoker:2017mds} and shown to be related to the conformal fixed points of field theories derived from  taking a conformal limit of $(p,q)$ 5-brane webs. Various aspects of these solutions have been studied recently, see e.g.~\cite{Gutowski:2017edr,Gutperle:2017tjo,DHoker:2017zwj,Kaidi:2017bmd,Gutperle:2018vdd,Gutperle:2018wuk,Bergman:2018hin,Fluder:2018chf,Kaidi:2018zkx,Lozano:2018pcp,Hong:2018amk,Malek:2018zcz,Choi:2018fdc}.

The goal of the present note is to relate the form of the local IIB solutions  found in \cite{Apruzzi:2014qva} to the ones found in \cite{DHoker:2016ujz} and determine the exact map between the two. In addition we analyze the regularity conditions and the map for global regular solutions. The structure of the note is as follows. In sections 2 and 3 we briefly review the two supergravity solutions of \cite{Apruzzi:2014qva} and \cite{DHoker:2016ujz}  respectively. In section 4 we determine the  exact map between these two solutions, and illustrate the relation with some explicit examples. In section 5, we look at how the global regular  solution of \cite{DHoker:2017mds} is mapped into framework of \cite{Apruzzi:2014qva}. We conclude with a discussion in section 6.

\section{Review of AFPRT solutions}
\label{sec-2}

Here we outline the solution in \cite{Apruzzi:2014qva} by Apruzzi, Fazzi, Passias, Rosa, and Tomasiello (AFPRT).
The spacetime takes the form of AdS$_6 \times S^2 \times \Sigma_2$ and the supergravity fields depend on the two dimensional space  $\Sigma_2$ through four quantities $(x, \alpha, A, \phi)$, two of which are actually independent and can be used to parameterize $\Sigma_2$.
Following \cite{Apruzzi:2014qva}  we take $(x, \alpha)$ to be independent and $A = A(x, \alpha)$ and $\phi = \phi(x, \alpha)$ to be dependent functions. Here  $e^A$ is a  warping function and $e^\phi$ is the dilaton.
These quantities satisfy two partial differential equations,
\begin{align} \label{eq-2-pdes}
\dd{ \qty[ \frac{e^{4A-\phi}}{x} \cot\alpha \dd{(e^{2A}\cos\alpha)}+ \frac{1}{3x} e^{2A}\sqrt{1-x^2} \dd{(e^{4A-\phi}\sqrt{1-x^2}\sin\alpha)}] } = 0 \no \\
3 \sin(2\alpha) \dd{A} \wedge \dd{\phi} = \dd{\alpha} \wedge \qty\Big[ 6 \dd{A} + \sin^2\alpha\qty( - \dd{(x^2)} - 2(x^2+5)\dd{A} + (1+2x^2)\dd{\phi} )] 
\end{align}
The metric in the string frame is
\( \dd{s}_\text{S}^2 = \frac{\cos \alpha}{\sin^2 \alpha} \frac{\dd{q}^2}{q} + \frac{1}{9}q (1-x^2) \frac{\sin^2 \alpha}{\cos \alpha} \qty[ \frac{1}{x^2} \qty( \frac{\dd{p}}{p} + 3 \cot^2 \alpha \frac{\dd{q}}{q})^2 + \dd{s}^2_{S^2} ] + e^{2A} \dd{s}^2_{\text{AdS}_6} \label{eq-2-metric} \)
where $p,q$ are quantities defined by
\begin{align}
q &= e^{2A} \cos \alpha \no \\
p &= e^{4A - \phi} \sin \alpha \sqrt{1-x^2}
\end{align}
The one-form field strength $F_1$ is 
\( F_1 = s_1 s_2 \frac{e^{-\phi}}{6x \cos\alpha} \qty[ \frac{12 \dd{A}}{\sin\alpha} + 4e^{-A}(x^2-1) \dd{(e^A\sin\alpha)} + e^{2\phi} \sin\alpha \dd{\qty(e^{-2\phi}(1+2x^2))} ] \label{eq-2-oneform} \)
and the three-form NS-NS and R-R field strengths, $H_3$ and  $F_3$, are
\begin{align}\label{eq-2-threeforms} 
H_3 &= s_1 \frac{1}{9x} e^{2A}\sqrt{1-x^2} \sin\alpha \qty[- \frac{6\dd{A}}{\sin\alpha} + 2e^{-A}(1+x^2)\dd{(e^A \sin\alpha)} + \sin\alpha \dd{(\phi + x^2)}] \wedge \text{vol}_{S^2} \no \\
F_3 &= s_2 \frac{e^{2A-\phi}}{54} \sqrt{1-x^2} \frac{\sin^2\alpha}{\cos\alpha} \qty[ \frac{36 \dd{A}}{\sin\alpha}+ 4e^{-A}(x^2 - 7)\dd{(e^A \sin\alpha)} + e^{2\phi} \sin\alpha \dd{\qty(e^{-2\phi}(1+2x^2))} ] \wedge \text{vol}_{S^2} 
\end{align}
where $s_1$ and $s_2$ are $\pm$ signs and $\text{vol}_{S^2}$ denotes the volume form of $S^2$ with unit radius.
The self-dual five-form field strength $F_5$  vanishes in this background.
These field strengths satisfy the Bianchi identities
\begin{align} \label{eq-2-bianchi} 
0 &= \dd{F_1} \no \\
0 &= \dd{F_3} - H_3 \wedge F_1 \no \\
0 &= \dd{H_3}
\end{align}
The signs $s_1, s_2$ depend on the specific supergravity solution, which we discuss later in this note.

\section{Review of DGKU solutions}
\label{sec-3}

Here we outline the solution in \cite{DHoker:2016ujz} by D'Hoker, Gutperle, Karch, and Uhlemann (DGKU).
The spacetime takes the form of AdS$_6 \times S^2 \times \Sigma_2$, where $\Sigma_2$ is a Riemann surface parametrized by complex coordinates $w, \bar{w}$.
The supergravity fields depend only on $\Sigma_2$ through two holomorphic functions $\mathcal{A}_\pm(w)$.  
For completeness we present the following quantities  which are useful to express the supergravity fields in a concise form.
We use the notation $\partial \equiv \partial_w$ and $\bar{\partial} \equiv \partial_{\bar{w}}$.
\begin{align}\label{eq-3-basicdefs}
\kappa_\pm &= \partial \mathcal{A}_\pm \no \\
\kappa^2 &= - |\kappa_+|^2 + |\kappa_-|^2  \no \\
\partial \mathcal{B} &= \mathcal{A}_+ \partial \mathcal{A}_- - \mathcal{A}_- \partial \mathcal{A}_+ \no \\
\mathcal{G} &= |\mathcal{A}_+|^2 - |\mathcal{A}_-|^2 + \mathcal{B} + \bar{\mathcal{B}} \no \\
\mathcal{D} = \qty(\frac{1+R}{1-R})^2 &= 1 + \frac{2 |\partial \mathcal{G}|^2}{3 \kappa^2 \mathcal{G}} 
\end{align}
The metric in the Einstein frame is
\(\dd{s}_\text{E}^2= f_6^2 \dd{s}^2_{\text{AdS}_6} + f_2^2 \dd{s}^2_{S^2} + 4 \rho^2 |\dd{w} |^2 \label{eq-3-metric}\)
where the metric factors are
\begin{align}\label{eq-3-metriccomps} 
f_6^2 &= \frac{\kappa^2}{\rho^2} \sqrt{\mathcal{D}} \no \\
f_2^2 &= \frac{\kappa^2}{9\rho^2} \frac{1}{ \sqrt{\mathcal{D}}} \no \\
(\rho^2)^2 &= \frac{\kappa^4}{6 \mathcal{G}} \sqrt{\mathcal{D}} 
\end{align}
Note that to make contact with the parameterization in section \ref{sec-2}, the metric should be transformed into the string frame,
\begin{equation}
\dd{s} ^2_{\text{S}} = e^{\phi/ 2} \dd{s}_\text{E}^2
\end{equation}
Here the dilaton is normalized in the standard  fashion to $\tau = \chi + i e^{-\phi}$.
The solution  \cite{DHoker:2016ujz} utilizes an $SU(1,1)/U(1)$ parametrization of the complex scalar field in terms of $B$, which is related to the axion-dilaton field via
\( B = \frac{1 + i\tau}{1 - i\tau} \)
and is given by \cite{Hong:2018amk} in terms of the defined quantities as
\( B = \frac{\mathcal{S} + \mathcal{T}/\sqrt{\mathcal{D}}}{\bar{\mathcal{S}} - \bar{\mathcal{T}}/\sqrt{\mathcal{D}}} \)
where for notational convenience we introduced the quantities
\begin{align} \label{eq-3-STdef} 
\mathcal{S} &= -\mathcal{A}_+ + \bar{\mathcal{A}}_- \no \\
\mathcal{T} &= \frac{\kappa_+ \bar{\partial} \mathcal{G} + \bar{\kappa}_- \partial \mathcal{G}}{\kappa^2}
\end{align}
This gives expressions for the axion $\chi$ and the dilaton $e^\phi$,
\begin{align}
e^{\phi} &= -\frac{ (\mathcal{S} + \bar{\mathcal{S}})^2 - (\mathcal{T} - \bar{\mathcal{T}})^2/\mathcal{D}}{2(\mathcal{S} \bar{\mathcal{T}} + \bar{\mathcal{S}}\mathcal{T})/\sqrt{\mathcal{D}} } \no \\
\chi &= -i \, \frac{ (\mathcal{S}^2 - \bar{\mathcal{S}}^2) - (\mathcal{T}^2- \bar{\mathcal{T}}^2)/\mathcal{D}}{ (\mathcal{S} + \bar{\mathcal{S}})^2 - (\mathcal{T} - \bar{\mathcal{T}})^2/\mathcal{D}}
\end{align}
If we also define
\( \mathcal{U}_\pm = (\kappa_+ \pm \kappa_-) \bar{\partial} \mathcal{G} \)
then noting the relations
\begin{align} \label{eq-3-Udef}
\mathcal{U}_- + \bar{\mathcal{U}}_- &= \kappa^2 (\mathcal{S} + \bar{\mathcal{S}}) & \mathcal{U}_- - \bar{\mathcal{U}}_- &= \kappa^2 (\mathcal{T} - \bar{\mathcal{T}}) \no \\
\mathcal{U}_+ + \bar{\mathcal{U}}_+ &= \kappa^2 (\mathcal{T} + \bar{\mathcal{T}}) & \mathcal{U}_+ -  \bar{\mathcal{U}}_+ &= \kappa^2 (\mathcal{S} - \bar{\mathcal{S}})
 \end{align}
we have yet another expression for the axion and dilaton
\begin{align} \label{eq-3-dilaton}
e^\phi &= - \frac{(\Re \mathcal{U}_-)^2 + (\Im \mathcal{U}_-)^2/\mathcal{D}}{ (\Re \mathcal{U}_- \Re \mathcal{U}_+ + \Im \mathcal{U}_- \Im \mathcal{U}_+)/\sqrt{\mathcal{D}}} \no \\
&= \frac{(\Re \mathcal{U}_-)^2 + (\Im \mathcal{U}_-)^2/\mathcal{D}}{ |\partial \mathcal{G}|^2 \kappa^2/\sqrt{\mathcal{D}}}  \\
\chi &= \frac{ \Re \mathcal{U}_- \Im \mathcal{U}_+ - \Im \mathcal{U}_- \Re \mathcal{U}_+/\mathcal{D}}{(\Re \mathcal{U}_-)^2 + (\Im \mathcal{U}_-)^2/\mathcal{D}} 
\label{eq-3-axion}
\end{align}
The one-form field strength $F_1$ is given in terms of the axion $\chi$ by
\( F_1 = \dd{\chi} \)
The complex two-form potential $\mathcal{C}_2$ is given by
\( \mathcal{C}_2 = \frac{2i}{9}\qty[ \frac{\mathcal{T}}{\mathcal{D}} -3 (\mathcal{A}_+ + \bar{\mathcal{A}}_-) ] \text{vol}_{S^2} \)
This can be written in terms of the real two-form potentials $C_2$ and $B_2$,
\( \mathcal{C}_2 = B_2 + i C_2 \) 
where now
\begin{align}
B_2 &= - \frac{1}{9i} \qty[ \frac{\mathcal{T} - \bar{\mathcal{T}}}{\mathcal{D}} - 3 (\mathcal{A}_+ + \bar{\mathcal{A}}_- - \bar{\mathcal{A}}_+ - \mathcal{A}_-  ) ] \text{vol}_{S^2} \no \\
C_2 &= \phantom{-i} \frac{1}{9} \qty[ \frac{\mathcal{T} + \bar{\mathcal{T}}}{\mathcal{D}} - 3 (\mathcal{A}_+ + \bar{\mathcal{A}}_- + \bar{\mathcal{A}}_+ + \mathcal{A}_-  ) ] \text{vol}_{S^2}
\end{align}
This gives the R-R and NS-NS three-form field strengths $F_3$ and $H_3$
\begin{align} \label{eq-3-threeforms} 
F_3 &= \dd{C_2} - H_3 \chi \no \\
H_3 &= \dd{B_2} 
\end{align}
The self-dual five-form field strength $F_5$ vanishes.
These field strengths satisfy the same Bianchi identities \eqref{eq-2-bianchi} given previously.

\section{Mapping local solutions}
\label{sec-4}

Given these two different approaches to finding half-BPS solutions with $\text{AdS}_6$ factors in type IIB supergravity, our goal is to determine  how they are related.
We note that the difference in the parameterization of the solution lies in the two dimensional Riemann surface $\Sigma_2$.  The DGKU solution  uses a uniformized form with complex coordinates $w, \bar w$ whereas the AFPRT solution uses coordinates which are adapted to the pure spinor construction leading to the PDEs \eqref{eq-2-pdes}.

In order to relate the DGKU solution with the AFPRT solution the goal is to identify the four quantities $(x, \alpha, A, \phi)$ of AFPRT in terms of the coordinates $w, \bar{w}$ given the holomorphic data $\mathcal{A}_\pm(w)$ of DGKU. 
We use the fact that  the  four quantities  $(f_2^2, f_6^2, \chi, \phi)$ are scalars with respect to $\Sigma_2$ and hence  are independent of coordinate choices. Consequently they can be used to obtain a map between the two parameterizations.
We show in the following that the coordinates $x$ and $\alpha$ and the independent functions $A$ and $\phi$ can be expressed in terms of the holomorphic functions $\mathcal{A}_\pm(w)$ and that they satisfy the PDEs \eqref{eq-2-pdes}.

\subsection{Positivity}
\label{sec-4-0}

We start by slightly adapting the discussion of positivity  in \cite{DHoker:2016ujz}.
On the Riemann surface $\Sigma_2$, we consider solutions where the supergravity fields remain finite and the metric components are strictly positive:
\( 0 < f_2^2, f_6^2, \rho^2 \)
Then the definitions \eqref{eq-3-basicdefs} imply that $\kappa^2$, $\mathcal{G}$, and $\sqrt{\mathcal{D}}$ are non-zero, finite, and have the same sign.
From the definition of $\mathcal{D}$, we necessarily have $\mathcal{D} \geq 1$.
In taking the square-root we have a sign ambiguity, so without loss of generality we can take $\sqrt{\mathcal{D}} \geq 1$.
This gives us the equivalent constraints
\( 0 < \kappa^2 , \mathcal{G} < \infty \text{ on } \Sigma_2 \)

\subsection{Matching metric  factors}
\label{sec-4-1}

We can start by identifying the metrical factors  of $\dd{s}^2_{S^2}$ and $\dd{s}^2_{\text{AdS}_6}$ in the two string frame metrics \eqref{eq-2-metric} and \eqref{eq-3-metric}.
\begin{align} \label{eq-3-f6} 
f_2^2 e^{\phi/2} &= \frac{1}{9} e^{2A}(1-x^2) \sin^2 \alpha  \no \\
f_6^2e^{\phi/2} &= e^{2A} 
\end{align}
Then using the definitions in \eqref{eq-3-metriccomps} we have
\( \mathcal{D} = \frac{1}{(1-x^2) \sin^2\alpha}  \label{eq-4-D} \)
The dilaton $e^\phi$ is given explicitly in Eq.~\eqref{eq-3-dilaton}, and so is $e^A$ through Eq.~\eqref{eq-3-f6}.
Then including Eq.~\eqref{eq-4-D} above, we can express three quantities of AFPRT in terms of $w, \bar{w}$:
\[ (1-x^2)\sin^2\alpha \qcq e^A \qcq e^\phi \]
Only one more quantity needs to be matched.
If we equate the remaining portions of the two metrics, which correspond to $\dd{s}^2_\Sigma$, we can simplify to get
\(\cot^2 \alpha \qty( \frac{\dd{q}}{q})^2 + \frac{1}{9 \mathcal{D} x^2} \qty( \frac{\dd{p}}{p} + 3 \cot^2 \alpha \frac{\dd{q}}{q} )^2 = \frac{2\kappa^2}{3\mathcal{G}} \dd{w} \dd{\bar{w}} \label{eq-4-remainmetric} \)
We may also make the replacement $\cot^2\alpha = \mathcal{D}(1-x^2) - 1$.
This equation turns out to be not very helpful because it contains derivatives of $\alpha$ in $\dd{q}$. 
We can write the left-hand side in terms of $\alpha$, its first-order derivatives $\partial_w\alpha$ and $\partial_{\bar{w}}\alpha$, and other quantities we already know how to write in terms of $w, \bar{w}$.
Matching the differentials $\dd{w} \dd{w}$, $\dd{w}\dd{\bar{w}}$, and $\dd{\bar{w}} \dd{\bar{w}}$ on both sides will give first-order non-linear PDEs for $\alpha(w, \bar{w})$.
We will not attempt this approach as it turns out there is a more direct way match the last remaining quantity.

\subsection{Matching one-forms}
\label{sec-4-2}

The last remaining quantity can be matched using the one-form field strength.
In AFPRT, $F_1$ is given in Eq.~\eqref{eq-2-oneform}.
In DGKU, we have an expression for the axion $\chi$ in Eq.~\eqref{eq-3-axion}.
We can then simplify the equation $F_1 = \dd{\chi}$ to
\( 4 (3\mathcal{D}-1)(1-x^2)\dd{A}  - 2  (1 + 2x^2)\dd{\phi} + 2(1-x^2) \dd{ \ln \mathcal{D}} =  6s_1 s_2 x e^\phi \dd{\chi} (\mathcal{D}(1-x^2) - 1)^{1/2} \label{eq-4-oneformeq}\)
It is important to note that the $\dd{\alpha}$ dependence drops out.
Now apart from $x^2$, every quantity appearing in this equation (i.e.~$A$, $\phi$, $\mathcal{D}$, and $\chi$) can be written in terms of $w, \bar{w}$.
By making the replacement $\dd{} \to \partial_w$ and squaring of both sides of the equation, we obtain a complex-valued quadratic equation for $x^2$.
This quadratic equation is very complicated for general $\mathcal{A}_\pm$ functions, but will always have a real root with the surprisingly simple form
\( 1- x^2 = \frac{(\mathcal{S} + \bar{\mathcal{S}})^2 - (\mathcal{T} - \bar{\mathcal{T}})^2/\mathcal{D}}{(\mathcal{S} + \bar{\mathcal{S}})^2 - (\mathcal{T} - \bar{\mathcal{T}})^2} \)
This was arrived at by firstly taking explicit examples for $\mathcal{A}_\pm$ where the one-form equation \eqref{eq-4-oneformeq} was simple enough to be solved and guessing a general solution, and then verifying this solution algebraically for general $\mathcal{A}_\pm$ using Mathematica.
So far no insightful simplifications have been found; due to the simplicity of the solution despite the complexity of the quadratic equation, it is very possible that this conclusion can be obtained from simpler considerations.
Another useful form is given by
\( 1- x^2 = \frac{ (\Re \mathcal{U}_-)^2 + (\Im \mathcal{U}_-)^2 / \mathcal{D}}{ (\Re \mathcal{U}_-)^2 + (\Im \mathcal{U}_-)^2} \label{eq-4-x2sol}\)

\subsection{Explicit examples}
\label{sec-4-3}

Before continuing, let us verify  this mapping for  two established solutions which were discussed as examples in \cite{Apruzzi:2014qva}. \\

\textbf{Example 1 --}
The first example of a type IIB solution in \cite{Apruzzi:2014qva} is given by
\( e^A = \frac{c_1}{\cos^{1/6}\alpha} \qcq e^\phi = \frac{c_2}{\sin\alpha \cos^{2/3}\alpha} \qcq x = 0 \label{eq-4-ex1target}\)
for $0 < \alpha < \pi/2$.
As the only independent variable is $\alpha$, this solution is slightly degenerate.
The holomorphic data which reproduces this solution is given in \cite{DHoker:2016ujz}, and is slightly changed here for convenience.
\( \mathcal{A}_\pm = -\frac{a}{2} w^2 \pm ibw \qcq \mathcal{B}(w=0) = \frac{ab}{6} \)
Below we give some relevant derived quantities.
In this example, we use coordinates $w = (X+iY)/2$.

\begin{align} \label{eq-4-ex1}
\kappa^2 &= 2ab Y \no \\
\mathcal{G} &= \frac{1}{3}ab(1-Y^3) \no \\
\mathcal{D} &= \frac{1}{1-Y^3} \no \\
\mathcal{U}_- &= 2ab^2 Y^2 \no \\
e^\phi &= \frac{2b}{aY} \frac{1}{\sqrt{1 - Y^3}} \no \\
e^A &= \frac{\sqrt{2b}}{Y^{1/4}} \no \\
\chi &= \frac{a X}{2b} 
\end{align}
For positivity, we take $0 < Y < 1$.
The one-form equation \eqref{eq-4-oneformeq} simplifies to
\( \frac{ 9 x^2 Y\qty( 4 - 20x^2 +25 x^2 Y^3)}{(1-Y^3)^2} = 0 \)
and has a solution $x^2 = 0$.
This is consistent with our solution for $x^2$ in Eq.~\eqref{eq-4-x2sol} as $\mathcal{U}_-$ has zero imaginary part.
Then the equation for $\mathcal{D}$ in Eq.~\eqref{eq-4-D} implies the coordinate change
\( Y = \cos^{2/3}\alpha \)
Plugging this into our expressions for $e^\phi$ and $e^A$ in \eqref{eq-4-ex1} above gives exactly the desired solution \eqref{eq-4-ex1target}, with the identifications
\( c_1 = \sqrt{2b} \qcq c_2 = 2b/a \)
If we write out the metric starting from \eqref{eq-3-metric} with $w = (X + i\cos^{2/3}\alpha)/2$, we can match the metric in \cite[Eq.~(A.1)]{Lozano:2013oma} if we identify
\( \frac{1}{4}\hat{W}^2 \hat{L}^2 = \frac{2b}{9} \frac{1}{\cos^{1/3}\alpha} \qcq  \hat{\theta} = \alpha \qcq \hat{\phi}_3 = \frac{2b}{3} X \)
and the axion and dilaton if we identify
\( a = \frac{27}{16} \hat{L}^4 \hat{m}^{1/3} \qcq b = \frac{9}{8} \hat{L}^2 \hat{m}^{-1/3} \)
where quantities with $\hat{\phantom{a}}$ are those of \cite{Lozano:2013oma}.
Then the three-form field strengths $F_3$ and $H_3$ also match.
In the literature, this solution is obtained by a Hopf $T$-duality on the $\text{AdS}_6 \times S^4$ Brandhuber-Oz solution to massive IIA supergravity \cite{Brandhuber:1999np}. \\

\textbf{Example  2 --}
A second solution  given in \cite{Apruzzi:2014qva} is
\( e^A = \frac{c_1}{\cos^{1/6}\alpha} \qcq e^\phi = \frac{x c_2}{\sin^3\alpha \cos^{1/3}\alpha} \)
for $0 < \alpha < \pi/2$ and $0 < x < 1$.
The holomorphic data which reproduce this solution  is
\( \mathcal{A}_\pm = -\frac{a}{3} w^3 \mp i bw + i 18a \qcq \mathcal{B}(w=0) = 0 \)
Let us now try to rederive this AFPRT solution with this as our starting point.
Below we give some relevant derived quantities.
In this example, we use coordinates $w = X + 3iY$. 
\begin{align} \label{eq-4-ex2} 
\kappa^2 &= -24 ab XY \no \\
\mathcal{G} &= -144 ab X ( 1- Y^3) \no \\
\mathcal{D} &= \frac{X^2 Y + (1-Y^3)^2}{X^2 Y (1-Y^3)} \no \\
\mathcal{U}_- &= 144ab^2(XY^2 + i (1-Y^3))\no \\
e^{\phi} &= \frac{b}{6a} \frac{1}{\sqrt{Y(1-Y^3)(X^2Y + (1-Y^3)^2)}} \no \\
e^A &= \frac{\sqrt{12b}}{Y^{1/4}} \no \\
\chi &= \frac{a}{b} (-X^2 +15 Y^2 - 6Y^5)
\end{align}
For positivity, we take $0 < Y < 1$ and $X < 0$.
The one-form equation \eqref{eq-4-oneformeq} has two solutions for $x^2$: one complex-valued, which does not admit a simple expression, and one which coincides with our solution \eqref{eq-4-x2sol},
\begin{subequations} 
\( x^2 = \frac{(1-Y^3)^2}{X^2 Y + (1-Y^3)^2} \)
From the equation for $\mathcal{D}$ in Eq.~\eqref{eq-4-D}, we also have
\( \sin^2\alpha = 1 - Y^3 \)
\end{subequations}
Together, these imply the coordinate change
\begin{align}\label{eq-4-ex2map} 
X &= -\frac{\sin^2\alpha}{\cos^{1/3}\alpha} \frac{\sqrt{1-x^2}}{x} \no \\
Y &= \cos^{2/3}\alpha
\end{align}
This gives us the desired $e^\phi$ and $e^A$ with the identifications
\( c_1 = \sqrt{12b} \qcq c_2 = b/6a \)
We can also match the metric, dilaton, and field strengths in \cite[Eq.~(11)]{Lozano:2012au} if we identify
\( \frac{1}{4} \hat{W}^2 \hat{L}^2 = \frac{4b}{3} \frac{1}{\cos^{1/3}\alpha} \qcq \hat{\theta} = \alpha \qcq \hat{r} = \frac{4b}{3} \frac{\sin^2\alpha}{\cos^{1/3}\alpha} \frac{\sqrt{1-x^2}}{x} \no \)
\( a = \frac{3}{512} \hat{L}^6 \qcq b = \frac{3}{16} \hat{L}^2 \hat{m}^{-1/3} \)
where quantities with $\hat{\phantom{a}}$ are those of \cite{Lozano:2012au}.
Then the three-form field strengths $F_3$ and $H_3$ also match.
In the literature, this solution is obtained by a non-Abelian $T$-duality on the Brandhuber-Oz solution. 

\subsection{Summary of the relation}
\label{sec-4-4}
We have shown in this section
the four quantities $(x, \alpha, A, \phi)$ of AFPRT  can be expressed in terms of the holomorphic functions of DGKU  in the following way:
\begin{align} \label{eq-4-mappings} 
e^\phi &= \frac{(\Re \mathcal{U}_-)^2 + (\Im \mathcal{U}_-)^2/\mathcal{D}}{ |\partial \mathcal{G}|^2 \kappa^2/\sqrt{\mathcal{D}}} \no \\
e^{4A} &= \frac{(\Re \mathcal{U}_-)^2\mathcal{D} + (\Im \mathcal{U}_-)^2}{ |\partial \mathcal{G}|^2 \kappa^2/6\mathcal{G}} \no \\
1-x^2 &= \frac{ (\Re \mathcal{U}_-)^2 + (\Im \mathcal{U}_-)^2 / \mathcal{D}}{ (\Re \mathcal{U}_-)^2 + (\Im \mathcal{U}_-)^2} \no \\
\sin^2\alpha &= \frac{ (\Re \mathcal{U}_-)^2 + (\Im \mathcal{U}_-)^2}{ (\Re \mathcal{U}_-)^2 \mathcal{D} + (\Im \mathcal{U}_-)^2 }
\end{align}
We have verified that the map holds for two previously known solutions related to $T$-duals of type IIA solutions. 
For general local solutions the algebra becomes very extensive. 
The following steps in verifying the map of DGKU  to  AFPRT  have been performed algebraically for general $\mathcal{A}_\pm$ using Mathematica:
\begin{enumerate}
\item \emph{Match the remaining parts of the metric corresponding to $\dd{s}^2_\Sigma$.}
Much of the work has already been done in \eqref{eq-4-remainmetric}, but now we are able to take derivatives on the left-hand side with relative ease.
It also turns out quite nicely that $q ^2= (\mathcal{S} + \bar{\mathcal{S}})^2$ and $p = 6\mathcal{G}$.
\item \emph{Match the three-form field strengths.}
We can check that the equations for $H_3$ and $F_3$ in the AFPRT solution match those of the DGKU solution.
As our map only involves $x^2$ and $\sin^2\alpha$, it does not distinguish between the signs of $x$ or $\cos \alpha$.
These signs are fixed by the sign convention
\( s_1 = - \sign(x) \sign (\Im \mathcal{U}_-) \qcq s_2 = -\sign(\cos\alpha) \sign (\Re \mathcal{U}_-) \label{eq-4-sign} \)
Then the AFPRT three-form fields strengths in Eqs.~\eqref{eq-2-threeforms} simplify to
\begin{align}
H_3 &= - \frac{2}{9 (\mathcal{D}-1) } \, \frac{ (\Re \mathcal{U}_-)^2 + (\Im \mathcal{U}_-)^2}{\kappa^2\Im \mathcal{U}_-} \times  \no \\
& \qquad \qty[ - \frac{6 \dd{A}}{\sin^2\alpha} + (1+x^2)(2 \dd{A} + \dd{ (\ln \sin^2\alpha)}) + \dd{\phi} + \dd{(x^2)} ] \wedge \text{vol}_{S^2} \no \\
F_3 &= - \frac{ \mathcal{G}\sin^2\alpha}{18} \, \frac{\kappa^2}{\Re \mathcal{U}_-} \times \no  \\
& \qquad \qty[ \frac{36 \dd{A}}{\sin^2\alpha} + 2 (x^2 - 7) (2 \dd{A} + \dd{(\ln \sin^2\alpha)}) -2 (1+2x^2) \dd{\phi} + 2 \dd{(x^2)} ]  \wedge \text{vol}_{S^2} 
\end{align}
These are equivalent to those of DGKU in Eqs.~\eqref{eq-3-threeforms}.
This means that in section \ref{sec-4-3}, for Example 1 we take $s_2 = -1$, and for Example 2 we take $s_1 = -1$ and $s_2 = +1$.
\item \emph{Check the two PDEs.}
We can simplify the first PDE of \eqref{eq-2-pdes} to
\( \dd{ \qty[ \frac{\mathcal{G} \mathcal{D} \Re \mathcal{U}_-}{\Im \mathcal{U}_-} \dd{\qty(\frac{\Re \mathcal{U}_-}{\kappa^2}) } + \frac{1}{3(\mathcal{D}-1)}  \frac{(\Re \mathcal{U}_-)^2 \mathcal{D} + (\Im \mathcal{U}_-)^2}{\kappa^2 \Im \mathcal{U}_-} \dd{ \mathcal{G}} ]} = 0 \)
while the second PDE has no significant simplification.
\end{enumerate}
In summary we have shown that the general local DGKU  solution can be mapped to the AFPRT parametrization and satisfies the PDEs \eqref{eq-2-pdes} as a consequence of the holomorphy of the $\mathcal{A}_\pm(w)$.

\section{Mapping global solutions}
\label{sec-5}

After constructing a map from the local DGKU to  AFPRT solutions, we now look at the global solutions constructed in  \cite{DHoker:2017mds}.
They constitute a class of solutions (i.e.~specified $\mathcal{A}_\pm$ functions) where $\Sigma$ is taken to be the upper half-plane of the $w$ complex plane, whose boundary is the real axis.
The $\partial \mathcal{A}_\pm$ have poles on the boundary. The geometry is completely regular everywhere, except for the location of the poles where the supergravity background becomes that of a $(p,q)$ five-brane.  The supergravity solution can be viewed as the conformal near horizon limit of a $(p,q)$ five-brane web and the poles are interpreted as the residues of the semi-infinite external five-branes framing the web. 
If we take $(x, \alpha)$ to be alternative coordinates of $\Sigma$, we can ask how these features are mapped over.
We will find that while these global solutions are represented by a single coordinate patch on the $w$ complex plane, mapping over to the $(x, \alpha)$-coordinates requires multiple coordinate patches in order to have single-valued supergravity fields.

For simplicity we define
\( 1/\sqrt{\mathcal{D}} = \sin\alpha \sqrt{1-x^2} \)
In particular, this means $\sin\alpha \geq 0$.
Then as $x^2 \leq 1$ by definition, the square
\( -1 \leq x \leq 1 \qq{and} 0 \leq \alpha \leq \pi\)
becomes a very natural coordinate system for $(x, \alpha)$.
We will adopt this coordinate system for this section.

\subsection{Boundary conditions}
\label{sec-5-1}

We are mainly interested in solutions where the AdS$_6$ factor governs the entire non-compact part of the geometry. 
Therefore, we will assume that $\Sigma$ is compact, with or without boundary. 
On a non-empty boundary $\partial \Sigma$, we enforce $f_2^2 = 0$ while keeping the other conditions the same.
Physically, this corresponds to shrinking the $S^2$ sphere closing off the geometry and forming  a regular three-cycle which carries the five brane charges. 
This is equivalent to the boundary conditions
\( \kappa^2 = \mathcal{G} = 0 \text{ and } 0 < \mathcal{G}/\kappa^2 < \infty \text{ on } \partial \Sigma \)
The $0 < \mathcal{G}/\kappa^2$ constraint is relaxed at isolated points on the boundary to allow for sufficiently mild singularities, such as at the poles corresponding to five-branes. 

We can now make some general remarks on the boundary $\partial \Sigma$.
As $|\partial \mathcal{G}|^2 \neq 0$ we have $1/\mathcal{D} = 0$.
From the definition of $\mathcal{U}_-$ in Eqs.~\eqref{eq-3-Udef}, we also have $\Re \mathcal{U}_- = 0$.
Therefore
\begin{align} 
& 1-x^2 && = &&\frac{ (\Re \mathcal{U}_-)^2 + (\Im \mathcal{U}_-)^2 / \mathcal{D}}{ (\Re \mathcal{U}_-)^2 + (\Im \mathcal{U}_-)^2} &&\longrightarrow && 0 \no \\
& \sin^2\alpha && = &&\frac{ (\Re \mathcal{U}_-)^2 + (\Im \mathcal{U}_-)^2}{ (\Re \mathcal{U}_-)^2 \mathcal{D} + (\Im \mathcal{U}_-)^2 } &&\longrightarrow && \frac{(\Im \mathcal{U}_-)^2}{ (\Re \mathcal{U}_-)^2 \mathcal{D} + (\Im \mathcal{U}_-)^2 }
\end{align}

Thus $x^2 = 1$ is fixed, but because $(\Re \mathcal{U}_-)^2 \mathcal{D} \sim \kappa^2/\mathcal{G}$ is non-zero and finite (away from the five-brane poles), $\sin^2\alpha$ can take generic values on the interval $[0, 1]$.
Therefore we can say that \textbf{the boundary $\partial \Sigma$ corresponds to (segments of) the $x = \pm 1$ edges of the $(x, \alpha)$ square}.
Note that the boundary may not necessarily be mapped to the \emph{entire} edge $0 \leq \alpha \leq \pi$, but can map to just a segment of the edge.

\subsection{Example: non-Abelian $T$-dual}
\label{sec-5-2}

\begin{figure}
\centering
\begin{subfigure}[b]{0.4\textwidth}
	\includegraphics[width=\textwidth]{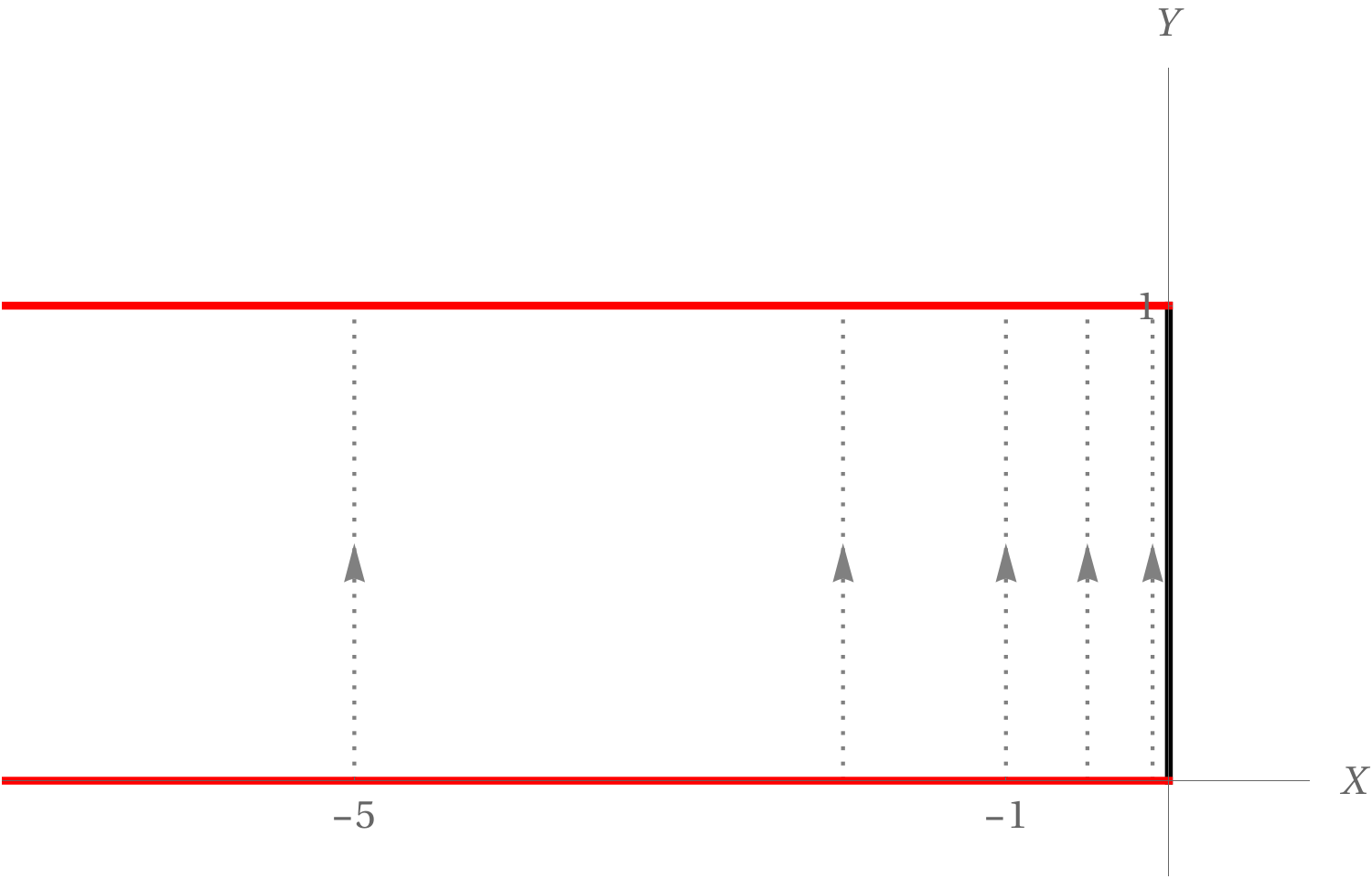}
	\caption{$\Sigma$ in $w$-coordinates}
	\label{fig-5-ex2-a}
\end{subfigure}
\hspace{1in}
\begin{subfigure}[b]{0.3\textwidth}
	\includegraphics[width=\textwidth]{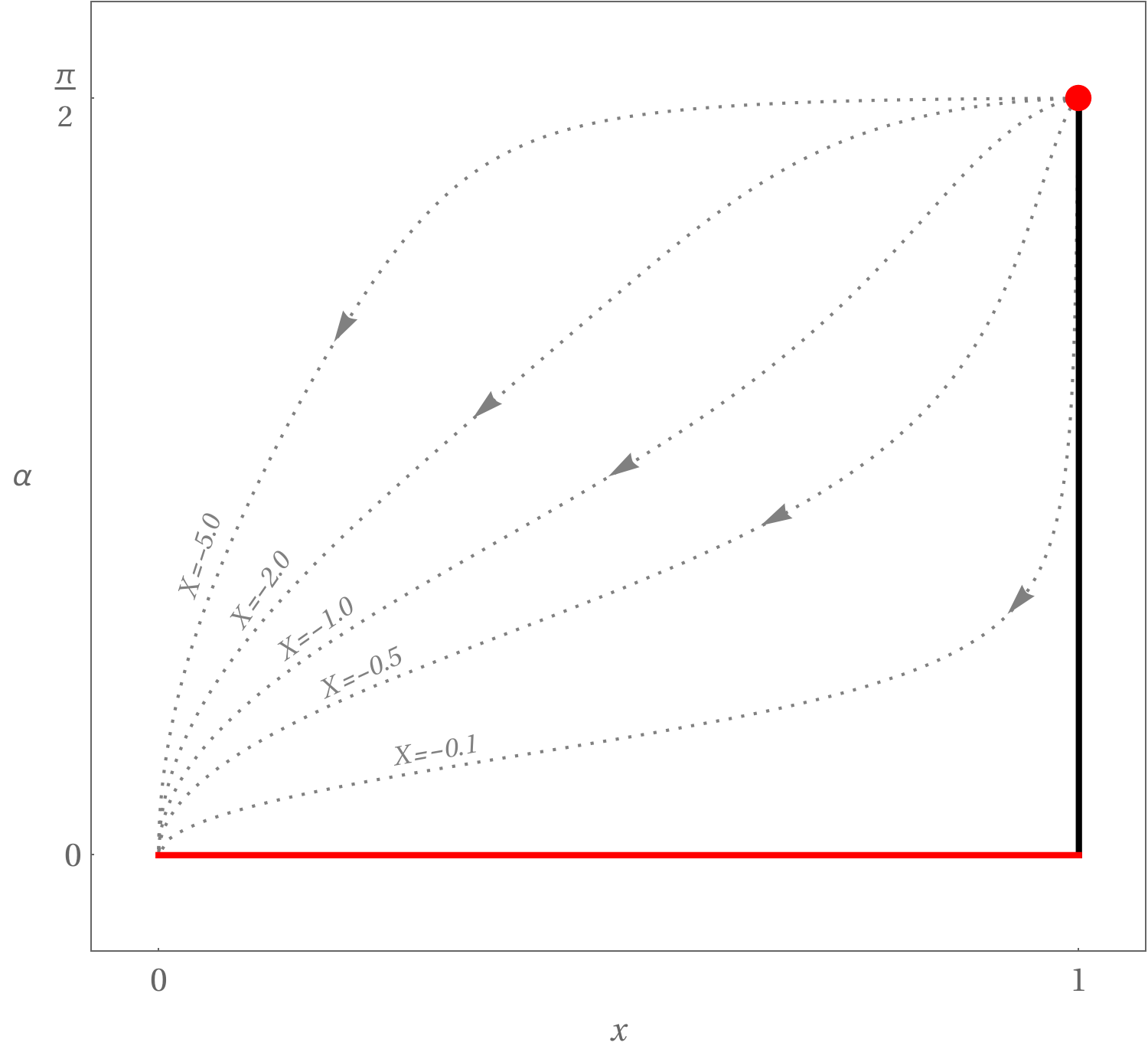}
	\caption{$\Sigma$ in $(x,\alpha)$-coordinates}
	\label{fig-5-ex2-b}
\end{subfigure}
\caption{Showing $\Sigma$ in different in coordinate systems.} \label{fig-5-ex2}
\end{figure}

As a warm-up, let us return to the second example of section \ref{sec-4-3}, with $a=1/4$ and $b = 1/6$ for concreteness.
We will first identify the Riemann surface $\Sigma$ on the $w$ complex plane, and then see how this region maps into the $(x, \alpha)$ square.
Recall that $w = X + 3i Y$.
\begin{align}
\kappa^2 &= -XY \no \\
\mathcal{G} &= -6 X ( 1- Y^3)
\end{align}
We satisfy $0 < \kappa^2, \mathcal{G}$ on the semi-infinite strip $X< 0$ and $0 < Y < 1$, which we take to be $\Sigma$.
Additionally, $\kappa^2 = \mathcal{G} = 0$ on the line segment $X = 0$ and $0 \leq Y \leq 1$, which we take to be the boundary $\partial \Sigma$.
The semi-infinite lines at $Y=0$ and $Y = 1$ are then coordinate singularities, where various metric components blow up.
\begin{align} 
f_2^2 &= \frac{2}{3} X^2 Y^{3/4} (1-Y^3)^{5/4} (X^2 Y + (1-Y^3)^2)^{-3/4} \no \\
f_6^2 &= 6 Y^{-1/4} (1-Y^3)^{1/4} (X^2 Y + (1-Y^3)^2)^{1/4} \no \\
\rho^2 &= \frac{1}{6} Y^{3/4} (1-Y^3)^{-3/4} (X^2 Y + (1-Y^3)^2)^{1/4} 
\end{align}
The coordinate patch for $\Sigma$ on the $w$ complex plane is shown in figure~\ref{fig-5-ex2-a}.
The black line represents the boundary $\partial \Sigma$, and the red lines represent the coordinate singularities. 

The coordinate change, given in Eq.~\eqref{eq-4-ex2map}, maps this semi-infinite strip on the $w$ complex plane into the quadrant $[0, 1] \times [0, \pi/2]$ of the $(x, \alpha)$ square. 
Explicitly,
\begin{align}
x &= \sqrt{ \frac{(1-Y^3)^2}{X^2 Y + (1-Y^3)^2}} \no \\
\alpha &= \sin^{-1} \sqrt{1-Y^3} 
\end{align}
Features of this map are shown in figure~\ref{fig-5-ex2}.
The boundary maps onto the line segment $x = +1$ as expected.
We have also included contours for visual aid, represented by dotted gray lines.
On the left-hand side we have drawn some contours of constant $X$, and on the right-hand side we show their images in the $(x, \alpha)$-coordinates.

\subsection{Example: 3-pole global solution}
\label{sec-5-3}

Let us finally turn our attention to the global solutions. We summarize  the relevant details from \cite{DHoker:2017mds}  for the general case of $L$ poles, and then specialize to a concrete example of three poles. 

Take $\Sigma$ to be the upper half-plane of the $w$ complex plane, and $\partial \Sigma$ to be the real-axis.
Let $p_\ell \in \mathbb{R}$ be the locations of $L$ poles on the real-axis, and $s_n \in \mathbb{H}$ be the locations of $N$ zeros strictly in the upper half-plane, where $L = N + 2$.
Then let
\( \partial \mathcal{A}_\pm(w) = \sum_{\ell = 1}^L \frac{Z_\pm^\ell}{w - p_\ell} \)
where $Z_\pm^\ell$ for $\ell = 1, 2, \dotsc, L$ are constants defined by
\( Z_+^\ell = i C_0 \prod_{n=1}^N (p_\ell - s_n) \prod_{\substack{\ell' = 1 \\ \ell' \neq \ell}}^L \frac{1}{(p_\ell - p_{\ell'})} \qcq Z_-^\ell = - \overline{Z_+^\ell} \)
and $C_0 \in \mathbb{C}$ is a complex-valued normalization constant.\footnote{In terms of the original paper, $i C_0 = \omega_0\lambda_0$ where $\bar{\omega}_0 = -\omega_0$ and $| \lambda_0| = 1$.}
If we integrate the expressions for $\partial \mathcal{A}_\pm(w)$, we have
\begin{subequations} \begin{align}
\mathcal{A}_+(w) &= \mathcal{A}^0 + \sum_{\ell=1}^L Z^\ell_+ \ln(w - p_\ell) \\
\mathcal{A}_-(w) &= -\overline{\mathcal{A}^0} + \sum_{\ell=1}^L Z^\ell_- \ln(w - p_\ell) 
\end{align} \end{subequations}
where $\mathcal{A}^0 \in \mathbb{C}$ is a constant satisfying the equation below for $k = 1, 2, \dotsc, L$.
\( \mathcal{A}^0 Z^k_- + \overline{\mathcal{A}^0} Z^k_+ + \sum_{\substack{\ell = 1 \\  \ell \neq k}}^L (Z^\ell_+ Z^k_- - Z^\ell_- Z^k_+) \ln |p_\ell - p_k| = 0 \label{eq-5-A0eq} \)
This makes $\kappa^2$ and $\mathcal{G}$ vanish on the real-axis, according to the usual definitions in \eqref{eq-3-basicdefs} after some consideration of branch cuts.
$\mathcal{G}$ contains dilogarithms, so any quantity containing it in undifferentiated form (such as $\mathcal{D}$) will not admit a simple form.
However, near a pole we can look at the asymptotic behavior.
Let us consider a pole $p_m$ and take the semi-circle $w = p_m + r e^{i\theta}$, where $0 \leq \theta \leq \pi$ and $0 < r \ll  |p_m - p_\ell|$ for all $\ell \neq m$.
Then we have the following relevant leading behaviors:
\begin{subequations} \begin{align}
\mathcal{D} &\approx \frac{|\ln r|}{3 \sin^2\theta} \\
\Re \mathcal{U}_- &\approx \kappa_m^2 (Z^m_+ - Z^m_-) \frac{|\ln r|}{r} \sin\theta \\
\Im  \mathcal{U}_- &\approx \kappa_m^2 (Z^m_+ - Z^m_-) \frac{|\ln r|}{r} \cos\theta
\end{align} \end{subequations}
where
\( \kappa_m^2 = -2i \sum_{\substack{\ell = 1 \\\ell \neq m}}^L \frac{ Z^m_+ Z^\ell_-  - Z^m_- Z^\ell_+ }{p_m - p_\ell} \)
so that near a pole as $r \to 0$
\begin{subequations} \begin{align}
& 1 - x^2 && = && \frac{ (\Re \mathcal{U}_-)^2 + (\Im \mathcal{U}_-)^2 / \mathcal{D}}{ (\Re \mathcal{U}_-)^2 + (\Im \mathcal{U}_-)^2} && \longrightarrow && \sin^2\theta &&&& \\
& \sin^2\alpha && = && \frac{(\Re \mathcal{U}_-)^2 + (\Im \mathcal{U}_-)^2}{(\Re \mathcal{U}_-)^2 \mathcal{D} + (\Im \mathcal{U}_-)^2} && \longrightarrow && \frac{3}{|\ln r|} && \longrightarrow && 0
\end{align} \end{subequations}
Therefore, small semi-circles around a pole on the $w$ complex plane map to lines of constant $\alpha$ on the $(x, \alpha)$ square, which approach either the $\alpha = 0$ or $\alpha = \pi$ edge as $r \to 0$.
Because $x$ is approximately $\pm \cos\theta$, these semi-circles necessarily map to the \emph{entire} line segment running between $-1 \leq x \leq 1$.
This can all be loosely summarized by saying \textbf{poles on the boundary $\partial \Sigma$ correspond to the $\alpha = 0, \pi$ edges of the $(x, \alpha)$ square}. 

For concreteness let us take the 3-pole solution, which is the simplest global solution with the fewest number of poles.
We pick the locations of the three poles,
\( p_1 = 1 \qcq p_2 = 0 \qcq p_3 = -1 \)
the location of the one zero,
\( s = \frac{1}{2}+ 2i \)
and the normalization constant,
\( C_0 = 1 \)
The relations \eqref{eq-5-A0eq} are solved by $\mathcal{A}^0 = i C_0 s \ln 2$. 

\begin{figure}
\centering
\begin{subfigure}[b]{0.4\textwidth}
	\includegraphics[width=\textwidth]{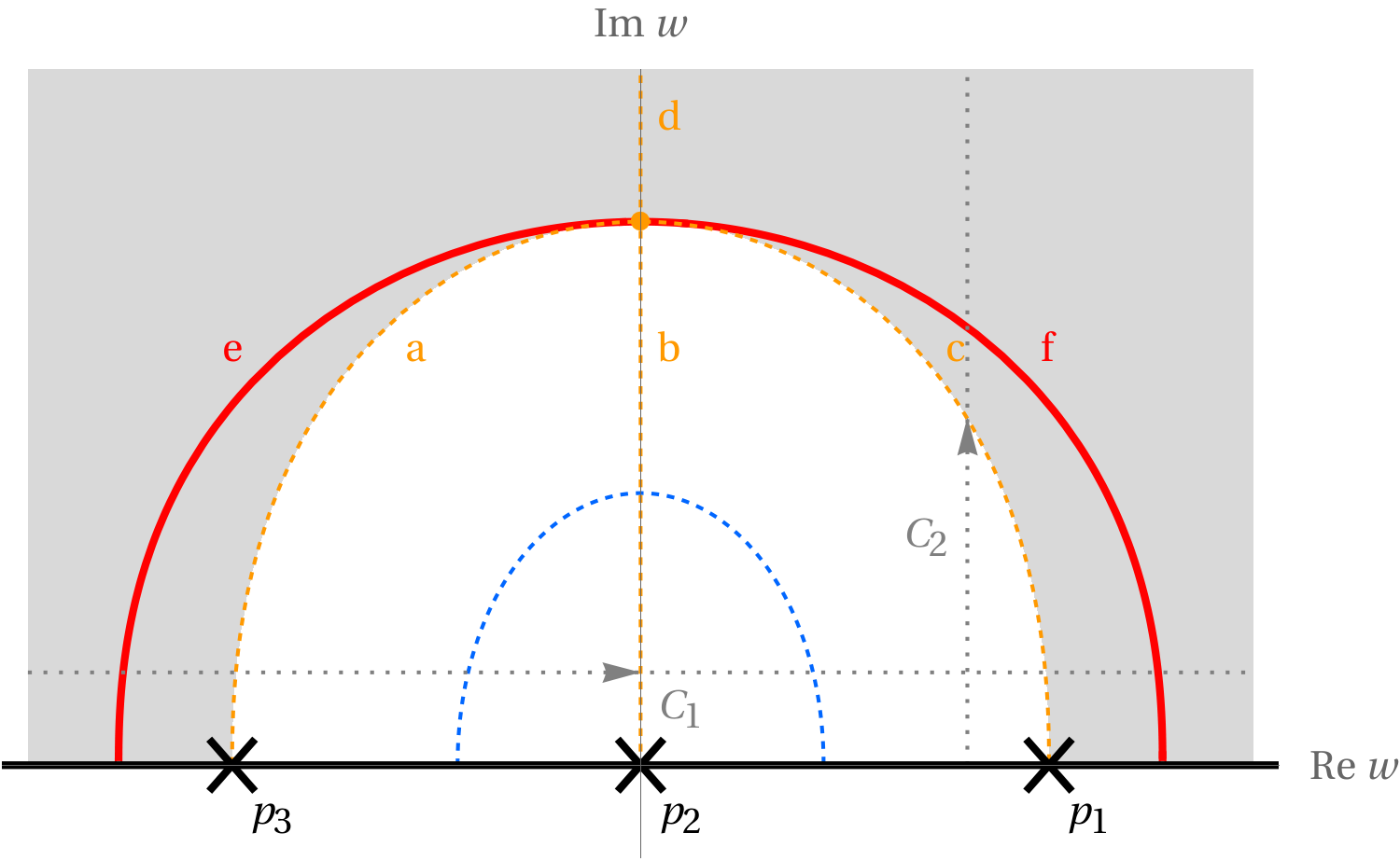}
	\vspace{-0.75cm}
	\caption{}
	\label{fig-5-3pole-a}
\end{subfigure}
\hspace{1in}
\begin{subfigure}[b]{0.3\textwidth}
	\includegraphics[width=\textwidth]{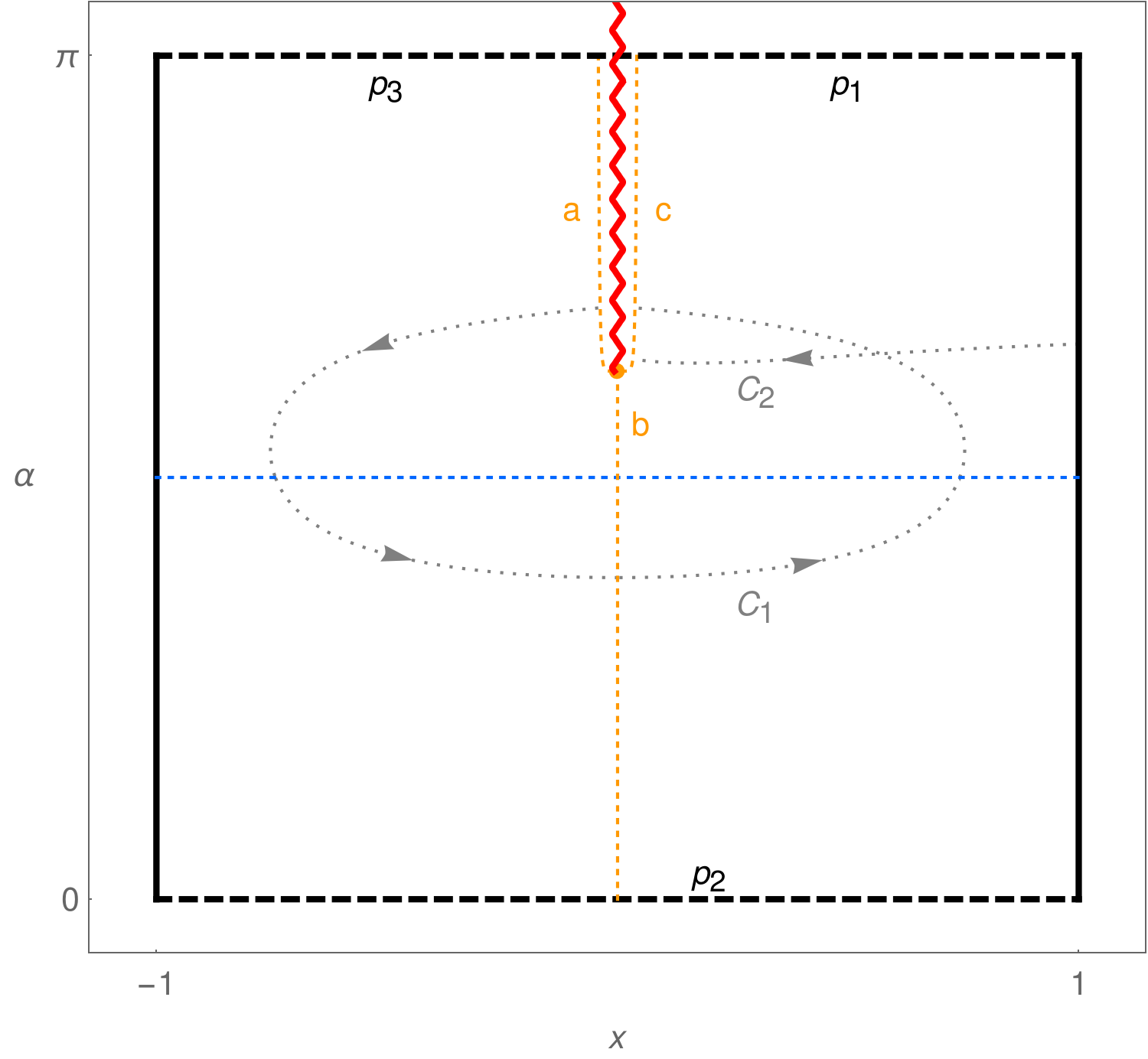}
	\vspace{-0.75cm}
	\caption{}
	\label{fig-5-3pole-b}
\end{subfigure}
\\
\vspace{0.5cm}
\begin{subfigure}[b]{0.4\textwidth}
	\includegraphics[width=\textwidth]{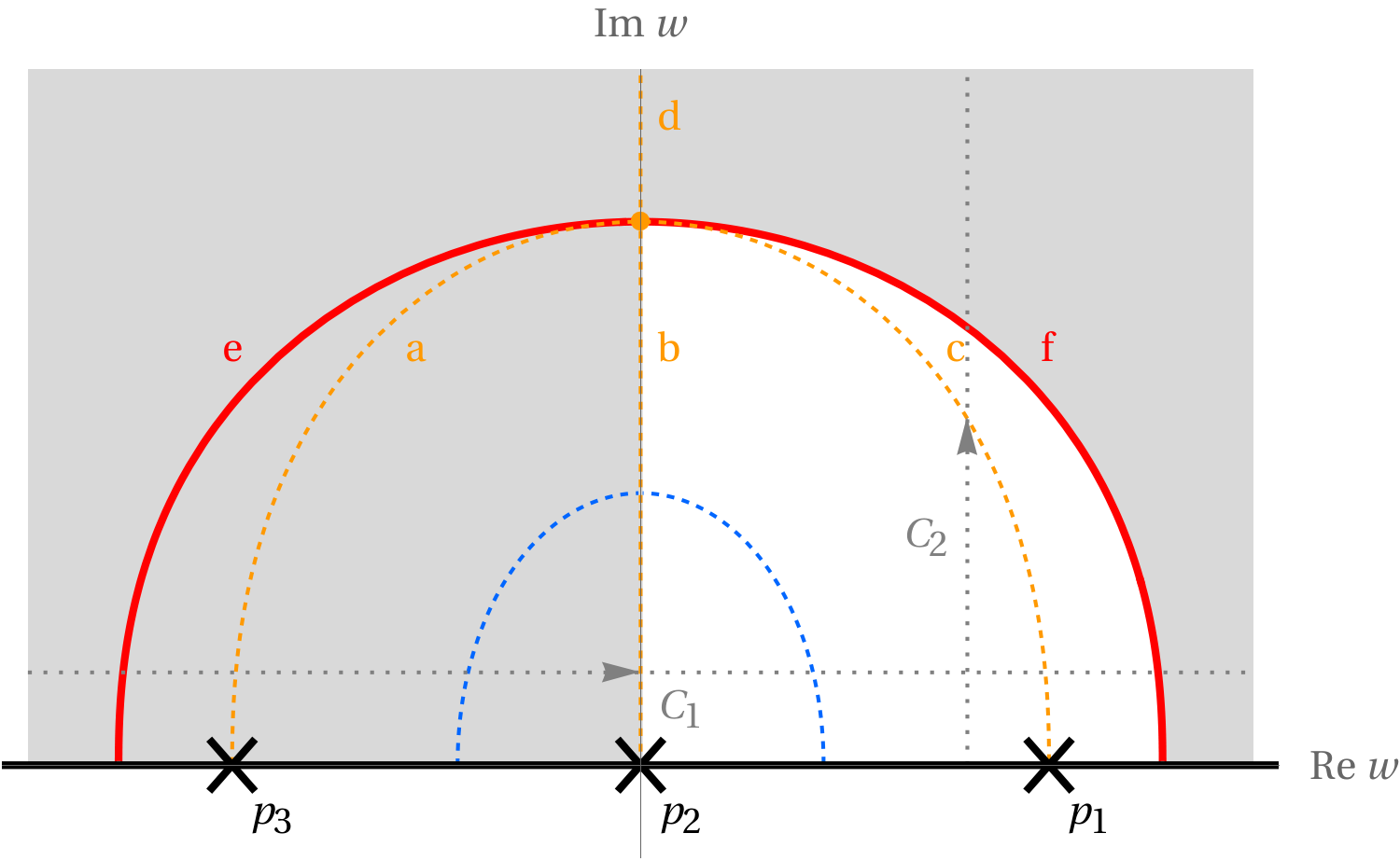}
	\vspace{-0.75cm}
	\caption{}
	\label{fig-5-3pole-c}
\end{subfigure}
\hspace{1in}
\begin{subfigure}[b]{0.3\textwidth}
	\includegraphics[width=\textwidth]{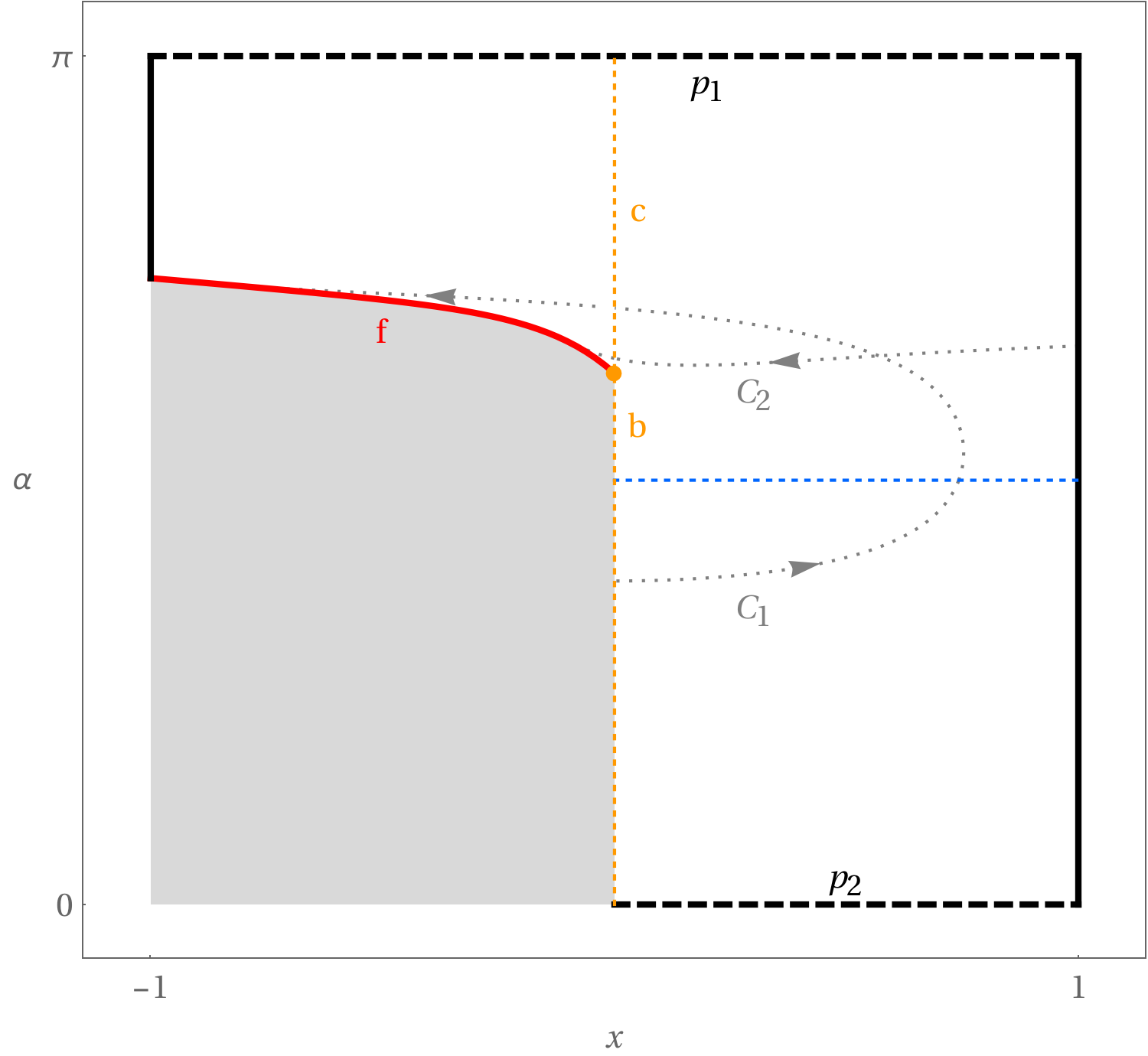}
	\vspace{-0.75cm}
	\caption{}
	\label{fig-5-3pole-d}
\end{subfigure}
\\
\vspace{0.5cm}
\begin{subfigure}[b]{0.4\textwidth}
	\includegraphics[width=\textwidth]{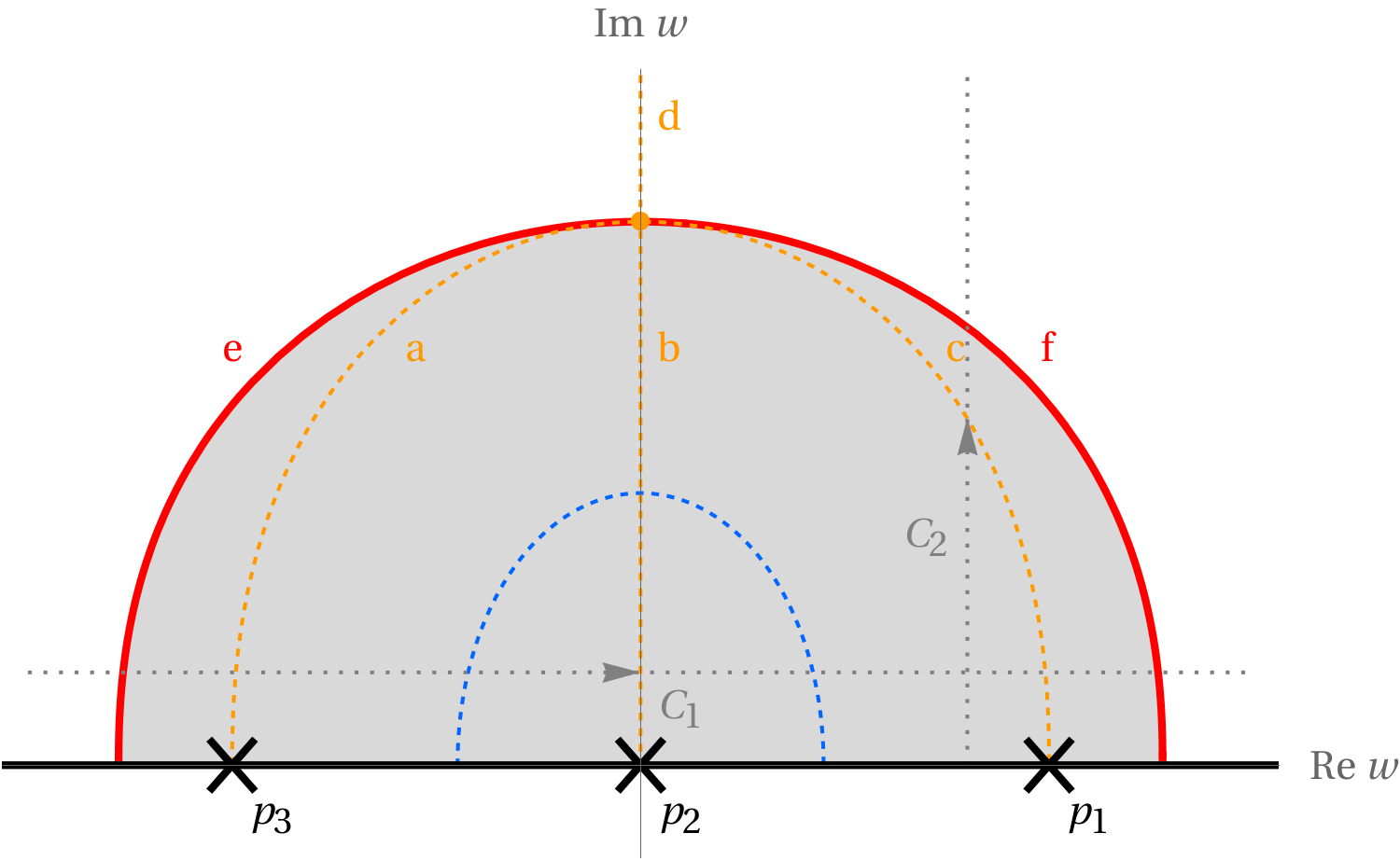}
	\vspace{-0.75cm}
	\caption{}
	\label{fig-5-3pole-e}
\end{subfigure}
\hspace{1in}
\begin{subfigure}[b]{0.3\textwidth}
	\includegraphics[width=\textwidth]{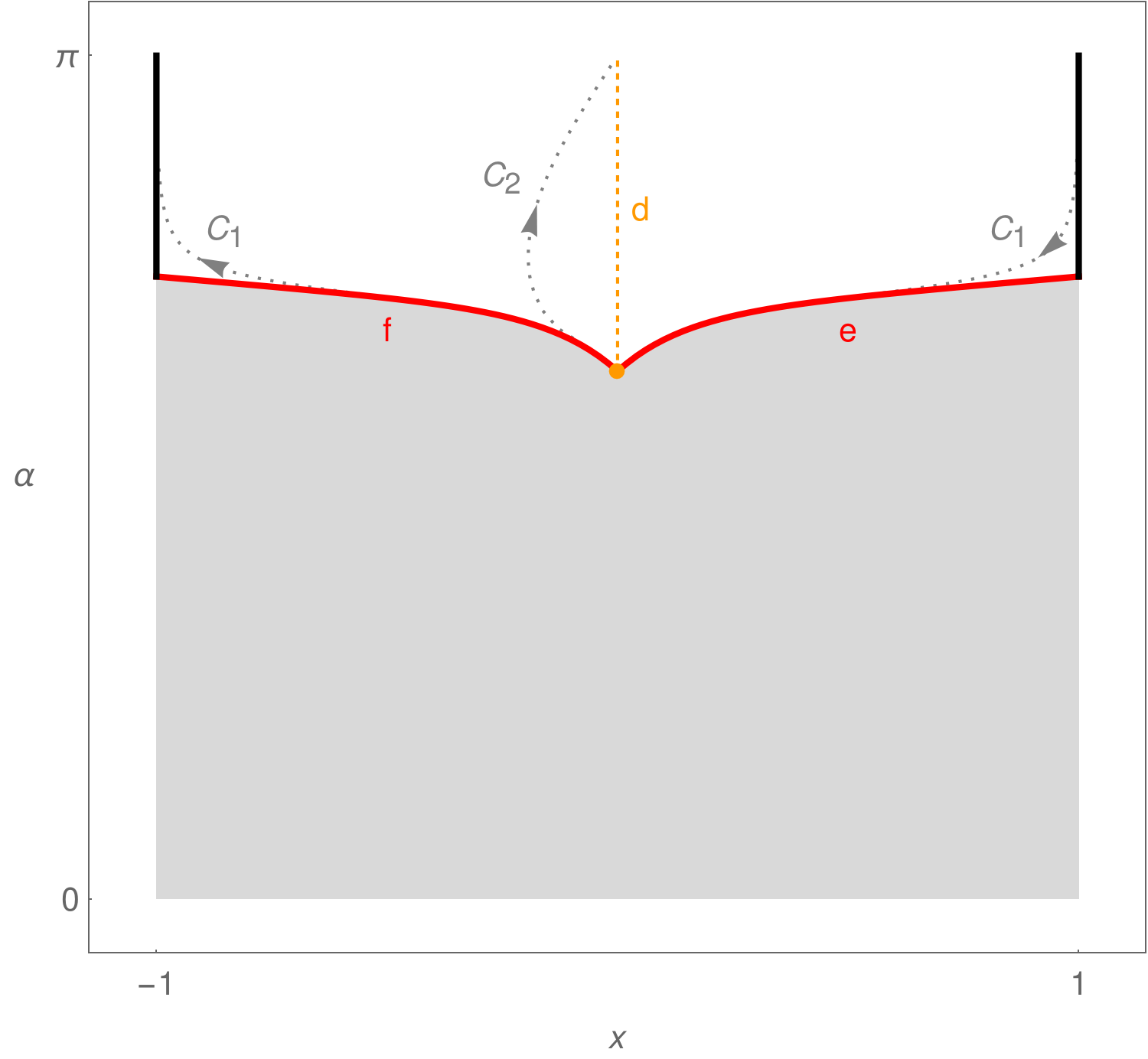}
	\vspace{-0.75cm}
	\caption{}
	\label{fig-5-3pole-f}
\end{subfigure}
\\
\vspace{0.5cm}
\begin{subfigure}[b]{0.4\textwidth}
	\includegraphics[width=\textwidth]{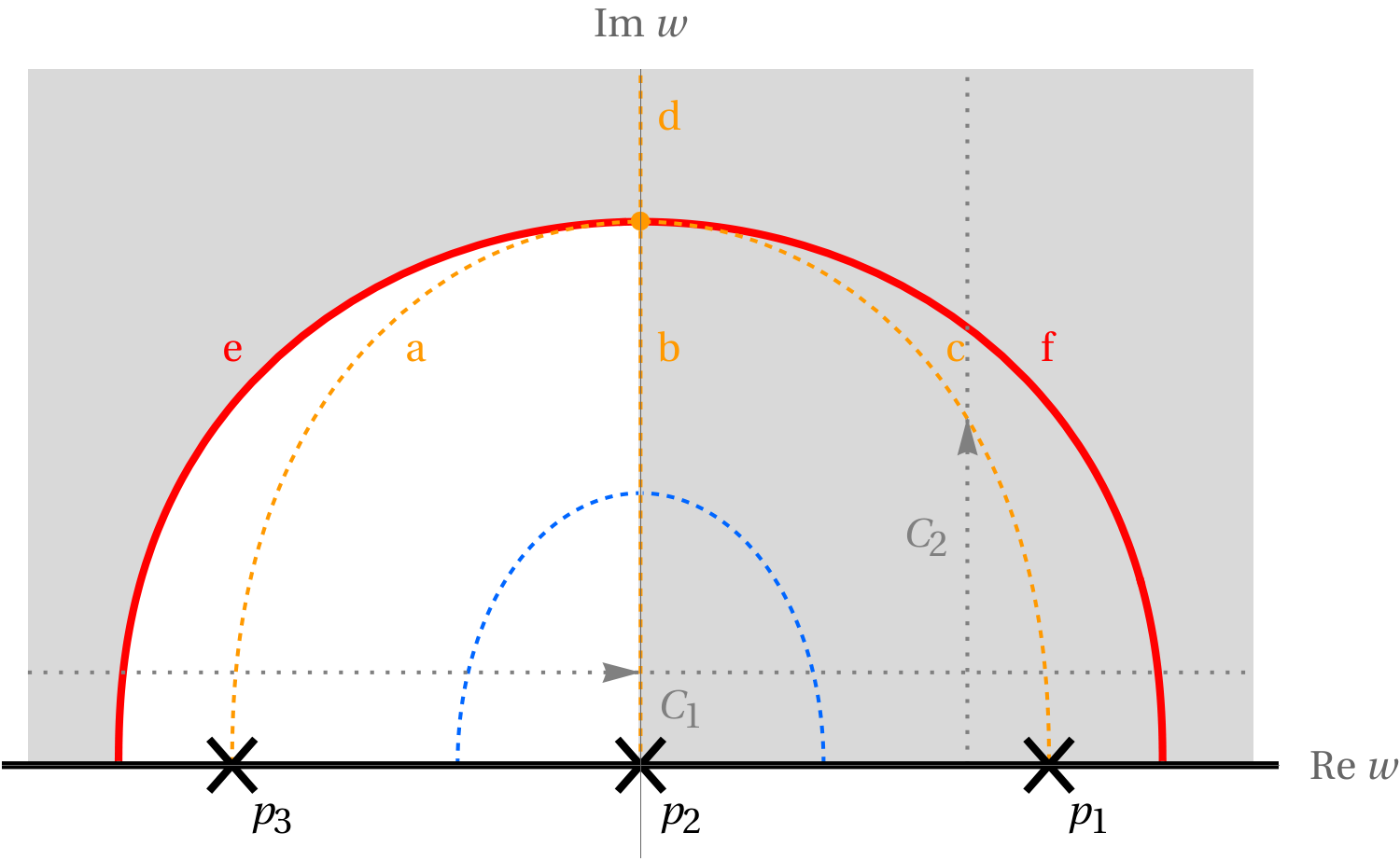}
	\vspace{-0.75cm}
	\caption{}
	\label{fig-5-3pole-g}
\end{subfigure}
\hspace{1in}
\begin{subfigure}[b]{0.3\textwidth}
	\includegraphics[width=\textwidth]{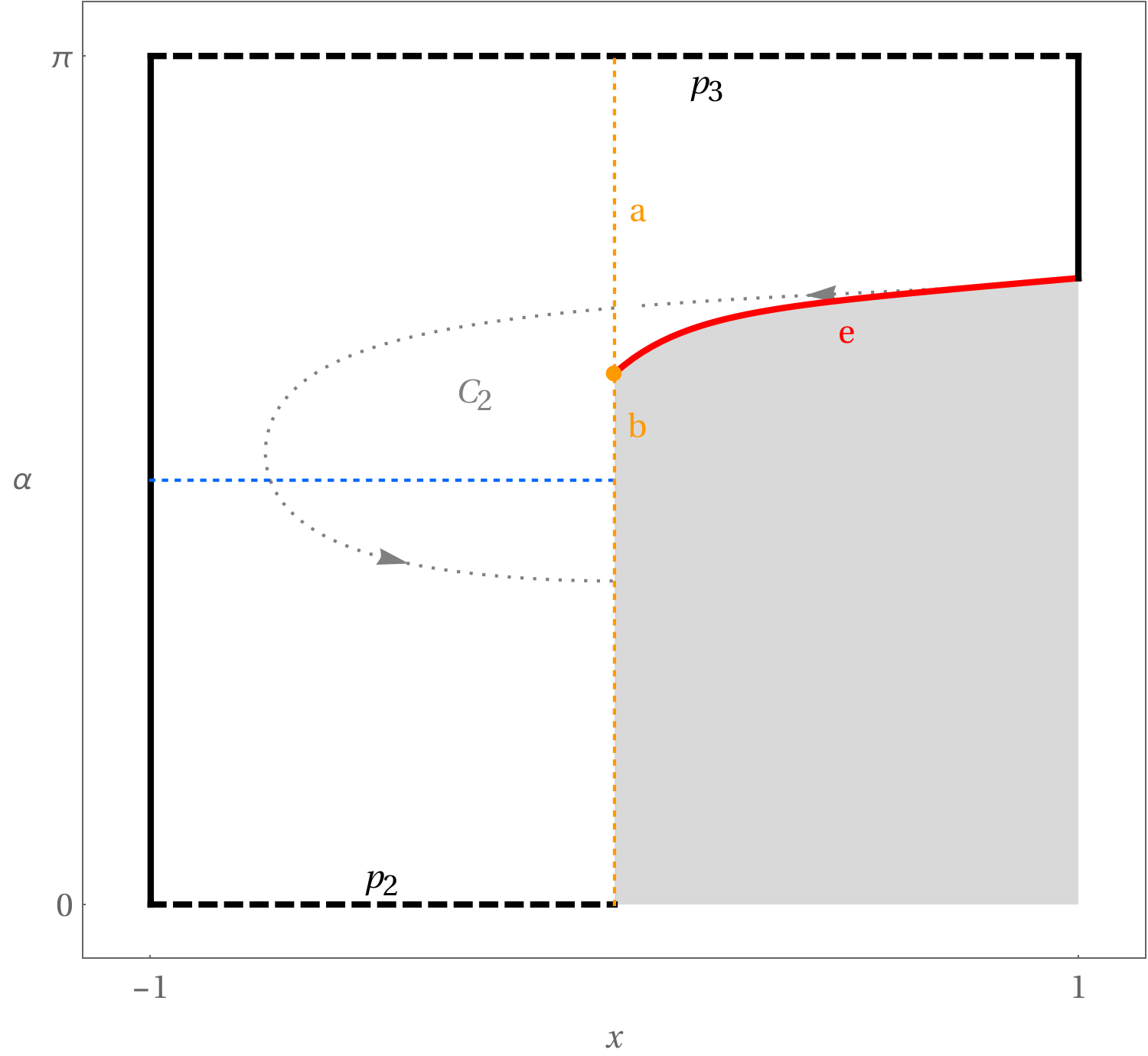}
	\vspace{-0.75cm}
	\caption{}
	\label{fig-5-3pole-h}
\end{subfigure}

\caption{Coordinate patches needed for the 3-pole solution.} \label{fig-5-3pole}
\end{figure}

This defines a coordinate change $(w, \bar{w}) \to (x, \alpha)$ from the upper half-plane into the square $[-1, 1] \times [0, \pi]$.
This map does not admit a simple form as it contains dilogarithms, but its general features are shown in figure~\ref{fig-5-3pole}.
The left-hand diagrams show an unshaded region on the $w$ complex plane, and the right-hand diagrams show the corresponding region on the $(x,\alpha)$ square.
Solid black lines represent the boundary $\partial \Sigma$.
``X'' marks on the $w$ complex plane represent locations of the five-brane poles, which map to black dashed lines on the $(x, \alpha)$ square.
Two contours $C_1, C_2$ are included for visual aid. 

There are three important considerations which make this map well-defined:
\begin{enumerate}
\item For convenience, we pick a map which obeys the sign convention \eqref{eq-4-sign} with $s_1 = s_2 = +1$.
On the $w$ complex plane, we have represented the curve where $\Im \mathcal{U}_- =0$ with an orange dotted line.
This curve is mapped to $x=0$.
The side of the curve where $\Im \mathcal{U}_- > 0$ gets mapped to the $x < 0$ side of the $(x, \alpha)$ square, and the side where $\Im \mathcal{U}_- < 0$ gets mapped to $x > 0$.
Similarly, the curve where $\Re \mathcal{U}_- = 0$ is represented with a blue dotted line.\footnote{$\Re \mathcal{U}_-$ also vanishes on the boundary $\partial \Sigma$, but we exclude this.}
\item In order to map the entire upper half-plane of the $w$ complex plane into $(x, \alpha)$-coordinates in a one-to-one manner, we need to introduce multiple coordinate patches.
This follows from a simple counting argument: each of the three poles needs to be mapped their own $\alpha = 0$ or $\alpha = \pi$ edge, only two of which exist on the square.
We can accommodate a one-to-one map at the expense of introducing multiple $(x, \alpha)$ squares and gluing them together.

For instance, the $p_2 = 0$ pole maps to the $\alpha = 0$ edge, and the boundary segments $p_3 < \Re w < p_2$ and $p_2 < \Re w < p_1$ map to the $x = -1$ and $x = 1$ edges, respectively.
The $p_1 = 1$ and $p_3 = -1$ poles then map to the $\alpha = \pi$ edge of two different $(x, \alpha)$ squares \ref{fig-5-3pole-d} and \ref{fig-5-3pole-h}, respectively. 
These two patches are glued together along the orange line ``b''.
Figure~\ref{fig-5-3pole-b} shows these two patches glued together at the expense of introducing a branch cut, represented by the red jagged line. 
\item The Jacobian $J$ of the map vanishes on the solid red line.
\( J \qq{=} \det \mqty( \partial x & \bar{\partial} x \\ \partial \alpha & \bar{\partial} \alpha ) \quad\propto\quad \qty[\partial\qty( \frac{\Re \mathcal{U}_-}{\Im \mathcal{U}_-}) \bar{\partial} \mathcal{D} - \bar{\partial}\qty( \frac{\Re \mathcal{U}_-}{\Im \mathcal{U}_-}) \partial \mathcal{D} ] \)
In the present example, if a contour on the $w$ complex plane passes through this line, the image of the contour in $(x, \alpha)$-coordinates will instead bounce off this line.
This means that we need an additional coordinate patch to maintain a one-to-one map.
For instance, consider the contour $C_2$ in $(x, \alpha)$-coordinates:  it starts on the coordinate patch \ref{fig-5-3pole-d}, hits the red line ``f'', and then bounces off onto another coordinate patch \ref{fig-5-3pole-f}.
\end{enumerate}

To summarize, $\Sigma$ is represented on the $w$ complex plane by a single coordinate patch, taken to be the upper half-plane.
When we map over to $(x,\alpha)$-coordinates, we need at least three coordinate patches to represent the whole $\Sigma$: \ref{fig-5-3pole-d}, \ref{fig-5-3pole-f}, and \ref{fig-5-3pole-h}.
\ref{fig-5-3pole-d} is glued to \ref{fig-5-3pole-h} along ``b'', \ref{fig-5-3pole-d} to \ref{fig-5-3pole-f} along ``f'', and \ref{fig-5-3pole-f} to \ref{fig-5-3pole-h} along ``e''.

\section{Discussion}
\label{sec-6}

In this note we have found an explicit map between the type IIB AdS$_6$ solutions formulated in \cite{Apruzzi:2014qva} and in \cite{DHoker:2016ujz}.
This mapping is given by a coordinate change $(w, \bar{w}) \to (x, \alpha)$ for the surface $\Sigma_2$ and was explicitly verified for two previously known examples. 
This result shows that the two solutions are indeed equivalent and that therefore the solutions of \cite{DHoker:2016ujz} are the most general type IIB solutions with an AdS$_6$ factor preserving sixteen supersymmetries.

Furthermore we mapped over the global solutions of \cite{DHoker:2017mds} and found that multiple coordinate patches in $(x, \alpha)$ were necessary in order to have single-valued solution.
This arose from a simple counting argument that each five-brane pole of the global solution needs to be mapped to its own horizontal edge of the $(x, \alpha)$ coordinate square, but global solutions have $\geq 3$ poles whereas each $(x, \alpha)$ square has 2 horizontal edges.
Thus an advantage of the complex coordinate parametrization of \cite{DHoker:2016ujz} is that a global solution can be represented in a single coordinate chart. Note that in   \cite{Apruzzi:2014qva} the four quantities $(A, \phi, x, \alpha)$ are initially on the treated same footing and subsequently $(x,\alpha)$ are chosen to be coordinates of the two dimensional space $\Sigma_2$. It is an interesting open question whether the global solutions can be formulated by making different coordinate choices.

In \cite{Kim:2015hya,Kim:2016rhs} the AFPRT solution was reduced on AdS$_6$ and an effective scalar coset theory was constructed.
The  symmetries of the coset can be used to  generate new solutions.  It would be interesting to investigate how the coset transformations act on the DGKU solutions using the mapping constructed in the present paper.
Another interesting direction would be to see how the alternative formulation in \cite{Apruzzi:2018cvq} can be related to the DKGU solutions. 

\section*{Acknowledgements}
We are grateful to Christoph Uhlemann for useful conversations.
The work of M.~G.~is supported in part by the National Science Foundation under grant PHY-16-19926.

\newpage


\providecommand{\href}[2]{#2}\begingroup\raggedright\endgroup

\end{document}